\documentclass{amsart}
\usepackage{xspace,colortbl}
\usepackage{amsmath,amsthm}

\usepackage{graphicx,here}
\usepackage[colorlinks]{hyperref}

\renewcommand{\(}{\left(}
\renewcommand{\)}{\right)}

\theoremstyle{definition}

\theoremstyle{remark}

\def\H{{\mathbb H}}

\newcommand{\one}{\leavevmode\hbox{\small1\normalsize\kern-.33em1}}

\numberwithin{equation}{section}

\begin{document}

\title{Time Quantization and $q$-deformations }

\author{Claudio Albanese}
\address{Claudio Albanese, Department of Mathematics, 100 St. George Street,
University of Toronto, M5S 3G3, Toronto, Canada}
\email{albanese@math.utoronto.ca}
\author{Stephan Lawi}
\address{Stephan Lawi, Department of Mathematics, 100 St. George Street,
University of Toronto, M5S 3G3, Toronto, Canada}
\email{slawi@math.utoronto.ca}

\date{August 26, 2003}

\begin{abstract}

We extend to quantum mechanics the technique of stochastic subordination,
by means of which one can express any semi-martingale as a time-changed
Brownian motion. As examples, we considered two versions of the
$q$-deformed Harmonic oscillator in both ordinary and imaginary time and
show how these various cases can be understood as different patterns of
time quantization rules.

\end{abstract}

\maketitle

\section{Introduction}

In a search to unravel the fabric of space at short distances, many
authors have explored variations on ordinary quantum mechanics based on
$q$-deformations of the canonical commutation relation, $q$ being a
parameter in the interval $(0,1)$ where 1 corresponds to the Bose limit,
see for instance \cite{ArikCoon}, \cite{Biedenharn}, \cite{Macfarlane}
Time quantization was considered also, see \cite{Caldirola},
\cite{RecamiFarias}. On an entirely different line of research,
probabilists developed the notion of stochastic time changes (also called
stochastic subordination) as a way of understanding jump processes,
see \cite{Bertoin}, \cite{Bochner}, \cite{Sato}. This
work gave rise to a representation of Levy processes, a family
of translation invariant jump processes, as subordinated Brownian motions
whereby the time change is uncorrelated to the underlying process. More
generally, Monroe proved that all semi-martingales can be represented as
time-changed Brownian motions, as long as one allows for the subordinator
to be correlated.

In this paper we bring together ideas from all these lines of research and
show that one can interpret $q$-deformations in terms of stochastic time
changes, albeit of a new type which is designed in such a way to preserve
quantum probability. We find that these representations provide new
insights in the notion of $q$-deformation and indicate an alternative,
physically intuitive path to understand short-scale deformations of
quantum field theory.

Stochastic subordination is a procedure to construct a stochastic
process from another by means of a stochastic time change.
If $X_t$ is a stochastic process, the subordinated
process $\tilde X_t$ is defined as follows:
\begin{equation}
    {\tilde X}_t = X_{T_t}
\end{equation}
where $T_t$ is a monotonously non decreasing process. The process $T_t$ is
called Bochner subordinator in case its increments $T_{t+\Delta t} - T_t$
are independent and their distribution depends only on the time elapsed
$\Delta t$ but not on $t$. Such time-homogeneity property makes Bochner
subordinator potentially appealing also for basic quantum physics.
Under such hypothesis, if the process $X_t$ is stationary and
Markov and $G$ is its generator, then also the process $\tilde X_t$
is a stationary Markov process. One can show under mild conditions
that the generator  $\tilde G$ of $\tilde X_t$
can be expressed as a function $\tilde G = \phi(G)$. It is useful
to sketch a quick proof in order to work out a quantum extension.

Suppose that the generator $G$ admits a complete set of eigenfunctions
$f_n$ with eigenvalues $\lambda_n$. (The argument below can
readily be extended to the case where $G$ has also continuous
spectrum). We have that
\begin{equation}
\rho_t(x) = \sum_{n=0}^\infty a_n e^{-\lambda_n t} f_n(x)
\end{equation}
where $\rho_0(x)$ is a fixed initial condition. If one instead
evolves the initial state according to the $\tilde X_t$
dynamics the distribution $\tilde \rho_t(x)$ is given by
\begin{equation}
\tilde \rho_t(x) = \sum_{n=0}^\infty a_n \left[ \int e^{-\lambda_n s}
\mu_t(ds) \right] f_n(x).
\end{equation}
Hence, there ought to be a function $\phi(\lambda)$ such that
\begin{equation}\label{eq:BochnerBernstein}
\int e^{-\lambda s} \mu_t(ds) = e^{-t \phi(\lambda)},
\end{equation}
an equation that needs to hold true for all times $t$ and all values of
$\lambda \ge 0$. This equation constrains the form of the functions
$\phi(\lambda)$ as not all functions admit a one-parameter family of
positive measures $\mu_t(ds)$ such as the equation above is satisfied. The
measures $\mu_t(ds)$, if well-defined, are called the renewal measures
\cite{Bertoin}.

Notice that Bochner subordination also applies to general
positivity preserving contraction semigroups $R_t$, as the
subordinated semigroup $\tilde R_t$ can be consistently
defined as follows:
\begin{equation}
\tilde R_t \rho_0 = \int R_s\rho_0\ \mu_t(ds).
\end{equation}

In the quantum mechanics case the problem is different as conservation of
quantum probability requires that the dynamics be defined by one-parameter
groups of unitary transformations. If $\mathbb H$ is a quantum Hamiltonian
and ${\mathbb U}_t = e^{-it{\mathbb H}}$ is the corresponding dynamics,
the subordinated dynamics is defined as follows:
\begin{equation}
\tilde {\mathbb U}_t \; = \; \int \; {\mathbb U}_s \;\chi_t(ds).
\end{equation}
Consistency with quantum probability conservation requires that
$\chi_t(ds)$ be a complex valued measure such that
\begin{equation}
\int e^{-i \lambda s} \chi_t(ds) = e^{- i t \phi(\lambda)}
\end{equation}
for some real valued function $\phi(\lambda)$. This condition restricts
the form of the measures $\chi_t(ds)$. As extensions to the renewal
measures, we shall call $\chi_t(ds)$ the {\it quantum renewal measures.}

In the following we consider several examples using
two versions of the $q$-deformed harmonic oscillator
as example. A first example is provided
by the one-mode harmonic oscillator Hamiltonian
\begin{equation}
{\mathbb H} = \frac{\hbar \omega}{ 2} (a^\dagger_q a_q + a_q a^\dagger_q)
\end{equation}
where the creation and annihilation operators $a_q$ and $a_q^\dagger$
satisfy the following relations
\begin{equation}
a_q a_q^\dagger - q a_q^\dagger a_q = 1.
\end{equation}
This deformation scheme was proposed by Arik and Coon in \cite{ArikCoon}
and leads to the energy spectrum
\begin{equation}
\epsilon_q(n) = \frac{\hbar \omega}{ 2}\ \frac{1- q^n }{ 1 - q}.
\end{equation}
This means that the $q$-deformed Hamiltonian operator $\tilde{\mathbb H}$
can be represented so that
\begin{equation}
\tilde{\mathbb H} = \phi({\mathbb H}) \equiv \frac{1 - q^{\mathbb H}}{ 1 -
q}.
\end{equation}
In the Euclidean picture, one considers the semigroup $e^{-t \tilde
{\mathbb H}}$, which can be represented as follows:
\begin{equation}
e^{-t \tilde {\mathbb H}} = e^{-\frac{t}{ 1-q}} \sum_{n=0}^\infty
\frac{t^n}{(1-q)^n n!} e^{-n \lvert \ln q \lvert {\mathbb H}} \; = \; \int
\; e^{-s {\mathbb H}} \;\mu_t(ds)
\end{equation}
where
\begin{equation}
\mu_t(ds) = e^{-\frac{t}{\delta t}} \sum_{n=0}^\infty \frac{1}{ n!} \left(
\frac{t}{\delta t} \right)^n \delta(s - n \lvert \ln (1-\delta t) \lvert)
ds.
\end{equation}
Notice that $\mu_t(ds)$ is the distribution at time $t$ of a Poisson
process of characteristic time $\delta t$.

In the real-time, quantum picture, we have to account for phases required
to maintain quantum probability conservation, namely
\begin{equation}
e^{- i t \tilde {\mathbb H}} = e^{- \frac{i t}{1-q}} \sum_{n=0}^\infty
\frac{(i t)^n}{(1-q)^n n!} e^{- n \lvert \ln q \lvert {\mathbb H}} \; =
\;\int \; e^{- i s {\mathbb H}} \;\chi_t(ds)
\end{equation}
where the quantum renewal measure characterizing the quantum subordinator is
given by
\begin{equation}\label{eq:AC_chi}
\chi_t(ds) = \frac{e^{-\frac{i t}{\delta t}}}{2\pi i} \sum_{n=0}^\infty
\frac{1}{ n!} \left( \frac{i t }{ \delta t} \right)^n \frac{1}{s +i n
\lvert \ln (1-\delta t) \lvert + i0}\ ds.
\end{equation}
By re-using the terminology in \cite{Caldirola}, the location of the poles
of the quantum renewal measure will be referred to as ``chronons'', or
time quanta.

In summary, these expressions show that the Arik-Coon $q$-deformation
scheme for the harmonic oscillator is equivalent in the Euclidean picture
to a time quantization, whereby imaginary time proceeds according to a
Poisson process and has increments equal to discrete quanta of $\delta t =
1 - q$. In the real, quantum-mechanics time picture, a similar Poisson
distribution applies except that additional phases have to be added to the
expansion in such a way to ensure conservation of quantum probability. In
the latter representation, chronons are located at equally spaced
intervals along the imaginary axis.

Alternative deformation rules have been proposed by Biedenharn in
\cite{Biedenharn} and MacFarlane in \cite{Macfarlane}. According to this
scheme $q$ is a parameter which can be not only real, but can also take
values in the unit circle $S^1 = \{ e^{i\alpha}, \alpha \in [0,2\pi) \}$.
According to this scheme the creation and annihilation operators $a_q$ and
$a_q^\dagger$ satisfy the following relations
\begin{equation}
a_q a_q^\dagger - q a_q^\dagger a_q = q^{-n}
\end{equation}
and the energy spectrum is
\begin{equation}
\epsilon_q(n) = \frac{\hbar \omega}{ 2}\ \frac{q^{n} - q^{-{n}}}{q -
q^{-{1}}}.
\end{equation}
In the following, we shall indeed assume that $q = e^{i\alpha}$, so that
the $q$-deformed Hamiltonian operator $\tilde{\mathbb H}$ is thus
\begin{equation}
\tilde{\mathbb H} = \phi({\mathbb H}) \equiv \frac{q^{\H} - q^{-{\H}}}{ q
- q^{-{1}}} = \frac{\sin\alpha\H }{ \sin\alpha}.
\end{equation}

The function $\phi$ does not correspond to a Bochner
subordinator, as there is no one-parameter family of positive
measures such that relation (\ref{eq:BochnerBernstein}) holds . Hence in
the Euclidean picture, one cannot consistently construct a positivity
preserving semigroup. In real-time however, there is not such a
restriction on the measure and we find
\begin{equation}
e^{- i t \tilde {\mathbb H}} =\sum_{n=0}^\infty \(\frac{-t}{
2\sin{\alpha}}\)^n\ \sum_{k=0}^n\frac{(-1)^k}{k!(n-k)!}\
e^{i{\alpha}(n-2k)\H} = \int e^{-is\H}\chi_t(ds)
\end{equation}
where the measure characterizing the quantum subordinator is
given by
\begin{equation}\label{eq:BM_chi}
\chi_t(ds) =\frac{1}{2\pi i}\sum_{n=0}^\infty \(\frac{-t}{
2\sin{\alpha}}\)^n\ \sum_{k=0}^n\frac{(-1)^k}{k!(n-k)!}\
\frac{1}{s+(n-2k)\alpha+i0}\ ds.
\end{equation}

\begin{figure}[H]
\begin{center}
\begin{tabular}{l r}
\begin{minipage}{6.5cm}
\begin{center}
\includegraphics[width = 14.7cm]{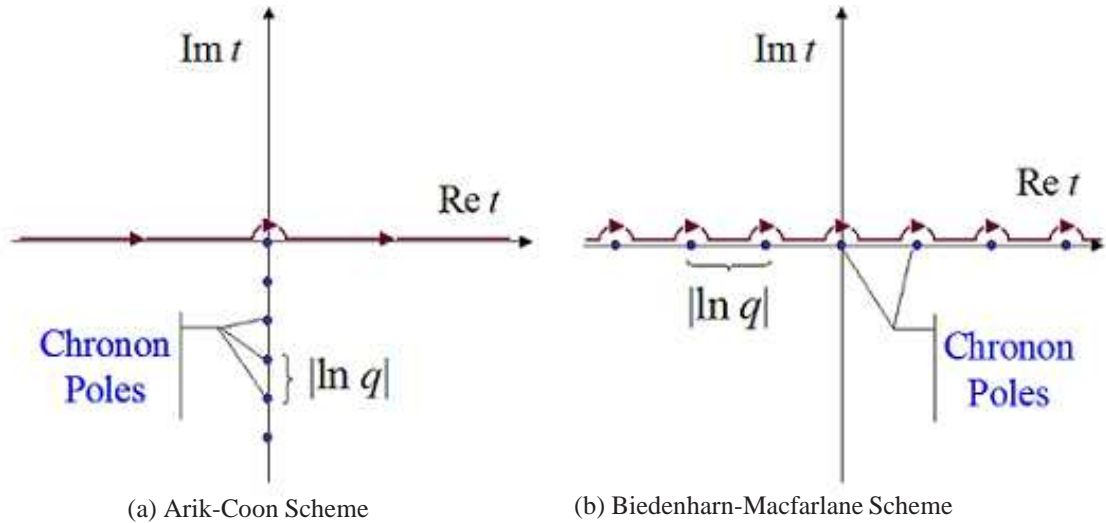}
\\ (a) Arik-Coon Scheme
\end{center}
\end{minipage}
&
\begin{minipage}{6.5cm}
\begin{center}
\vspace{6.5cm} (b) Biedenharn-Macfarlane Scheme
\end{center}
\end{minipage}
\end{tabular}
\vspace{0.5cm} \caption{Quantized Time Representations for Two
$q$-Deformation Schemes\label{fig:Quantization}}
\end{center}
\end{figure}

In this case, time is quantized along the real axis and the chronons, or
quantized time values, are poles of the measure $\chi_t(ds)$. The time
quanta, or the chronon increment, is given by $\alpha=\vert\ln q\vert$, as
in the Arik-Coon $q$-deformation scheme. It is instructive to compare both
$q$-deformation schemes and plot the distribution of the respective
chronons. Figure \ref{fig:Quantization} illustrates furthermore the
contour in the lower half plane, used to integrate the measures
$\chi_t(ds)$ in (\ref{eq:AC_chi}) and (\ref{eq:BM_chi}).

\section{Conclusions}

Stochastic subordination is one of the most common methodologies to deform
stochastic processes. In this paper, we introduce an analogue concept of
quantum subordination that is appropriate to deform quantum mechanical
systems. We show that this notion is broad enough to capture two examples
that received much attention in the literature, namely two versions of the
$q$-deformed Harmonic oscillator. In these two case, we show how quantum
subordination is defined by means of a quantum renewal function, an
analytic function with poles corresponding to time quanta or, to re-use
the terminology in \cite{Caldirola}, chronons. These concepts provide a
new angle to the notion of $q$-deformation and shed light on a physically
intuitive route to model short-scale deformations of quantum field theory.

\end{document}